# Linear frequency conversion via sudden merging of resonances in time-variant metasurfaces

Kanghee Lee[1,†], Jaehyeon Son[1,†], Jagang Park[1], Byungsoo Kang[1], Wonju Jeon[1], Fabian Rotermund[2] & Bumki Min[1,*]

[1]*Department of Mechanical Engineering, Korea Advanced Institute of Science and Technology (KAIST), Daejeon 305-751, Republic of Korea*

[2]*Department of Physics, Korea Advanced Institute of Science and Technology (KAIST), Daejeon 305-751, Republic of Korea*

[*] Correspondence to: bmin@kaist.ac.kr.

[†]These authors contributed equally to this work.

**Energy conversion in a physical system requires time-translation invariance breaking according to Noether's theorem[1,2]. Closely associated with this symmetry-conservation relation, the frequencies of electromagnetic waves are found to be converted as the waves propagate through a temporally varying medium[3-17]. Thus, effective temporal control of the medium, be it artificial or natural, through which the waves are propagating, lies at the heart of linear optical frequency conversion. Here, we propose rapidly time-variant metasurfaces as a frequency-conversion platform and experimentally demonstrate their efficacy at THz frequencies. The proposed metasurface is designed for the sudden merging of two distinct resonances into a single resonance upon ultrafast optical excitation. From this spectrally-engineered temporal boundary onward, the merged-resonance frequency component is radiated. In addition, temporal coherence of the two original resonating modes with respect to the abrupt temporal boundary is found to be strongly related to the amount of frequency conversion as well as the phase of the converted wave. Due to their design flexibility, time-variant metasurfaces may become on-demand frequency synthesizers for various frequency ranges.**

A few years before the invention of lasers and the subsequent experimental demonstration of the coherent nonlinear optical effect, which has since received much research attention, Morgenthaler contemplated an alternate route towards the conversion of light frequency via velocity modulation of the wave, i.e., temporal variations of permittivity and/or permeability of the media[3]. Although not much attention has been given to this seminal work relative to that given to nonlinear frequency conversion, continual theoretical efforts have been made to understand this linear frequency conversion under diverse physical conditions of time-varying media, such as spectral dispersion, spatial inhomogeneity, and the characteristics of temporal variation[4-7]. In line with these theoretical speculations, experimental verifications have followed in the field of plasma physics; notable examples are the observation of a frequency shift in a rapidly growing plasma from laser ionization of gases[8-10] and semiconductors[11], in which the electron density increased suddenly. However, it was not until a group of researchers proposed chip-scale dynamic photonic structures, including photonic crystals[12-15], micro-ring resonators[16], and waveguides[17] for the observation of wavelength conversion, that the phenomena became the subject of engineering efforts.

One lesson from previous studies is that it is necessary to make a large temporal change in the properties of media, through which light is transmitted, to maximize the frequency shift. Additionally, the introduction of a proper resonance structure enables miniaturization of the device, which in turn provides the additional benefit of maximizing temporal variation. For that reason, active resonant metamaterials[18-24], in which the resonating meta-atoms are hybridized with electrically or optically reconfigurable natural materials, can be thought of as an optimal platform that can be spatiotemporally rearranged. More specifically, a dramatic morphological transformation of the conducting resonance structures and their associated mode field profile changes will lead to the realization of ultimate linear frequency-conversion devices. Furthermore, the rapid temporal change in the thin time-variant metasurface will make it possible to clearly observe the coherent effect by limiting the spatiotemporal domain in which the new frequency component is generated.

To make this point more evident, the principle of frequency conversion in a time-variant metasurface is schematically illustrated in Fig. 1. The proposed time-variant metasurface consists of an array of two concentric split-ring resonators (SRRs) exhibiting two fundamental *LC* resonances, both of which are designed to be merged into a completely different resonance mode when the conductivity of the area between the two SRRs is suddenly increased. The time at which the conductivity abruptly changes is defined here as the temporal boundary. Depending on the time delay $t_d$ between the input pulse that drives the mode and this temporal boundary, the response of the time-variant metasurface can be categorized into three regimes of operation as depicted in Fig. 1. Here, a positive value of $t_d$ indicates that the input pulse reaches the metasurface before the time boundary arrives. When the temporal boundary is far behind (Fig. 1a) or ahead (Fig. 1c) of the input pulse, the modes driven by this input pulse are not disturbed; thus, the responses are those of a stationary metasurface. However, when the temporal boundary is coincident with the input pulse, more interesting time-varying phenomena can be observed because the operation of the metasurface cannot be simply described with these steady-state responses. While the original resonance modes still persist, a sudden merging of resonances, i.e., a spectrally-engineered rapid temporal variation of the metasurface, is induced, and forward from this temporal boundary the merged mode oscillating at $\omega_m$ (also satisfying the continuity of the surface current density) becomes the only fundamental eigenmode that can be sustained at the metasurface. Intuitively speaking, frequency conversion arises from this dynamic mode conversion process that occurs at the temporal boundary. Note that the proposed metasurface is just one of the possible implementations to achieve frequency conversion.

The proposed metasurface, shown in Fig. 2a, consists of an array of two concentric SRRs patterned on a semi-insulating GaAs substrate. The outer and inner SRRs in the unit cell were designed to exhibit fundamental resonance frequencies of $\omega_1/2\pi = 0.62$ THz and $\omega_2/2\pi = 1.24$ THz, respectively. Multiple dielectric layers were deposited on top of the patterned surface and etched selectively to expose only the area between the two SRRs (coloured in dark grey in Fig. 2a); the dielectric layers left behind by the etching were designed to reflect pump pulses centred at 800 nm, enabling the pump pulses to reach only the exposed area. As a result, the exposed GaAs area becomes conductive due to the pump-induced photocarrier generation. The conductivity change occurs within hundreds of femtoseconds and the changed conductivity is maintained over several tens of picoseconds[25,26]. Therefore, the temporal conductivity variation can be approximated with a step function. In all the subsequent experiments, incident THz and optical pump pulses propagate collinearly and enter the fabricated metasurface with a variable time delay, $t_d$. First, in order to investigate the steady-state response of the metasurface, measurements are taken under the condition that the optical pump pulse is delivered to the metasurface 4 ps prior to the arrival of the THz probe pulse (i.e., $t_d = -4$ ps). The combination of the step-like conductivity change of the GaAs and an adequate delay time guarantees that the THz pulse does not pass through the temporal discontinuity. Figure 2c shows the THz amplitude transmission spectra of the metasurface with a variation in optical pump fluence levels from 0 to 10 µJ/cm². As the pump fluence increases, the intrinsic semi-insulating area between the two SRRs becomes progressively more conductive, which morphologically alters the current-flow path within the unit cell and eventually merges the fundamental modes of the two SRRs into a new fundamental mode of a combined single SRR. The resonance frequency of the merged mode, $\omega_m$, falls between those of the original SRRs (0.92 THz at 10 µJ/cm²). Similar conditions were assumed in the Finite-Difference Time-Domain (FDTD) simulations, and the measured amplitude transmission spectra were well reproduced, as shown in Fig. 2d.

To understand the conversion mechanism more clearly, an infinitesimally thin conducting surface model, including frequency dispersion, is established for the metasurface (see Supplementary Information for details of the analysis). From the boundary conditions of Maxwell's equations, the electric field of the transmitted wave $\tilde{E}_t(\omega)$ can be determined from that of the incident wave $\tilde{E}_i(\omega)$ and the surface current density $\tilde{J}(\omega)$ as follows:

$$\tilde{E}_t(\omega) = \frac{2}{1+n}\left[\tilde{E}_i(\omega) - \frac{Z_0}{2}\tilde{J}(\omega)\right] \tag{1}$$

where $n$ is the refractive index of the dielectric substrate (half space) and $Z_0$ is the impedance of free space. The above equation is general in the sense that it is applicable to both time-invariant and time-variant systems. However, for time-variant systems, the second term plays a significant role because it contains the term related to the frequency generation as well as the conversion. In what follows, the response of a time-variant metasurface with a step-like temporal variation of a constituent material will be described with the steady-state responses of two time-invariant metasurfaces (prior and posterior to the temporal boundary) with an introduction of the electric field of a virtual incident wave, $E_v(t)$. The virtual electric field is defined in such a way that the steady-state response of the merged state to the *virtual* input field becomes equivalent to the steady-state response of the original state to the *real* input field (see Supplementary Information). Here, the steady-state response of the time-invariant metasurface refers to the transfer function that relates an input $\tilde{E}_i(\omega)$ to an output $\tilde{J}(\omega)$. For the original and merged states, the responses can be written as $\tilde{\theta}_{o,m}(\omega) = \frac{2\sigma_{o,m}(\omega)}{1+n+\sigma_{o,m}(\omega)Z_0}$, where $\sigma_{o,m}(\omega)$ are the effective surface conductivities of the metasurface before and after the pumping (see Supplementary Information). The surface current density spectrum for the step-like time-variant metasurface can then be written as

$$\tilde{J}(\omega) = \tilde{\theta}_o(\omega)\tilde{E}_i(\omega) + \tilde{\theta}_m(\omega)\int_{-\infty}^{\infty} u(t-t_d)[E_i(t) - E_v(t)]e^{i\omega t}\,dt \tag{2}$$

where $u(t)$ is a unit step function. The continuity of surface current densities at the temporal boundary is guaranteed by the definition of the virtual input field. The frequency conversion or generation is originated from the second term in Eq. 2 since the step function contains broad spectral components. Furthermore, it can be proven that $\tilde{\theta}_m(\omega)$ and $\tilde{\theta}_o(\omega)$ determine the efficiency of the frequency conversion.

To experimentally confirm the frequency conversion predicted by the theory described above, we first measured the amplitude transmission spectra of the metasurface as a function of the time delay $t_d$ between a sub-cycle THz input pulse and a pump pulse of 10 µJ/cm². The measured delay-dependent spectra are shown in Fig. 3a for time delays between -4.4 ps and 15.6 ps. The observed spectra show the presence of two original fundamental modes for an approximate range of $t_d > 10$ ps and the merged mode for $t_d < 0$ ps, corresponding to the steady-state responses of two time-invariant metasurfaces (and mathematically corresponding to the scenarios where $t_d$ approaches $\infty$ and $-\infty$, respectively, in Eq. 2). However, for an intermediate time delay of approximately $t_d = 2$ ps, where the original oscillating modes are disturbed by the temporal boundary, an amplitude transmission greater than unity is observed at $\omega_m$. To clearly visualize each of these situations, transmission snapshots at time delays of -4 ps (black), 2 ps (red), and 15 ps (blue) are shown in Fig. 3b. A plot of the transmission as a function of the time delay at the merged frequency (Fig. 3c) reveals another notable feature related to the coherence in the frequency conversion; more specifically, the delay-dependent oscillation in the transmission is clearly noticeable in the plot (see also the inset of Fig. 3c for the expanded view). The origin of the transmission oscillation is the delay-dependent interference between the incident and the converted field at the merged frequency. The calculation based on Eq. 2 is also plotted with a dotted line in Fig. 3c and exhibits a similar behaviour to the measured data. The sudden approximation is validated by the analytical plot calculated using the experimental transfer function, and the correlation between the measured and semi-analytical formulae is shown to be reasonably high. The measured data are also in excellent agreement with the full numerical simulations (see Supplementary Information for the validation of FDTD simulations under time-varying conditions and their calculated results).

To further investigate coherence in frequency conversion, a more detailed analysis was performed (see Supplementary Information); for a narrow band input pulse with a slowly varying envelope and a centre frequency of $\omega_i$, the following formula can be derived for the electric field of the transmitted wave (with an additional condition of $\omega \neq \omega_i$):

$$\tilde{E}_t(\omega) \approx -\frac{Z_0 \tilde{\theta}_m(\omega)}{1+n} \left| 1 - \frac{\tilde{\theta}_o(\omega_i)}{\tilde{\theta}_m(\omega_i)} \right| \frac{\alpha(t_d)}{2i} \left( \frac{e^{-i\phi - i(\omega + \omega_i)t_d}}{\omega + \omega_i} + \frac{e^{i\phi - i(\omega - \omega_i)t_d}}{\omega - \omega_i} \right) \quad (3)$$

where $\alpha$ describes an arbitrary slowly varying envelope and $\phi$ is the phase angle of $1 - \frac{\tilde{\theta}_o(\omega_i)}{\tilde{\theta}_m(\omega_i)}$. For the single band input, the converted field can be written as a weighted sum of two rotating complex amplitudes with an increasing time delay, $t_d$, in the complex plane, one rotating with $\omega + \omega_i$ and the other with $\omega - \omega_i$. In addition, the conversion efficiency is proportional to $\tilde{\theta}_m(\omega) \left| 1 - \frac{\tilde{\theta}_o(\omega_i)}{\tilde{\theta}_m(\omega_i)} \right|$, which is directly related to the effective conductivity engineering of the metasurface.

Figure 4 illustrates the theoretically calculated converted field for three different excitation schemes: (1) up-conversion from the lower fundamental to the merged frequencies (Fig. 4a); (2) down-conversion from the higher fundamental to the merged frequencies (Fig. 4b); and (3) superposition of the up-converted and down-converted signals from the lower and higher fundamentals to the merged frequencies (Fig. 4c). All the plots describe trajectories of the converted field on the complex plane with an increasing time delay. For the up-conversion, two complex amplitudes rotate counter clockwise with a dominant term rotating with $\omega - \omega_i$.

Pictorially, each of the two terms is plotted in Fig. 4a along with their sum; the plot resembles petals on a flower. For the down-conversion, however, two complex amplitudes rotate in the opposite directions (as shown in Fig. 4b), resulting in starfish-like plots on the complex plane. Figure 4c shows the case when the conversion occurs from both the higher and lower frequencies to the merged frequency. In the plots shown above, the input was assumed to be perfectly monochromatic. To simulate our experimental conditions, the trajectory in Fig. 4c is redrawn using values of $\alpha$ and $\phi$ corresponding to the experimental data to produce the plot shown in Fig. 4d. The trajectory appears complicated but has a predictable structured pattern. This type of coherence in the frequency conversion is clearly observable because of the sharp temporal boundary and infinitesimal thickness of the frequency conversion region. Furthermore, the coherence can be manipulated to a certain degree by changing the spectrally-engineered temporal boundary.

Experimentally, the corresponding data were measured and compared with the theoretical predictions; for these measurements, single- and dual-spectral-band multi-cycle THz input pulses were prepared using band-pass filters with transmission peak(s) centred around the original resonances (see the insets in Fig. 5 for the spectral content of each pulse; also see Supplementary Information for the preparation of the multi-cycle THz pulses). In this way, spectral components of input pulses corresponding to those of converted frequencies were carefully removed. Measured and calculated data with single-band input pulses of a centre frequency $\omega_{i1}$ are shown in Figs. 5a-e. The complex amplitude of the converted wave at the merged frequency $\omega_m$ is plotted in Fig. 5a as a function of the time delay along with its projections. The projected trajectory on the complex plane clearly shows a pattern similar to that theoretically calculated in Fig. 4a. This similarity between the measured and calculated data can also be confirmed in the evolutionary spectrum (i.e., the Fourier transform with respect to the time delay, see Fig. 5b); the two rotating frequency components of $\omega_m - \omega_{i1}$ and $\omega_m - \omega_{i2}$ are also prominent in the plot. In Figs. 5c and d, the complex amplitude magnitude for a frequency range of 0.8 - 1 THz and that at the merged frequency $\omega_m$ are plotted as a function of the time delay. The magnitude is directly related to the power conversion efficiency, $|E(\omega_m)|^2/|E(\omega_i)|^2$, which rises as high as $\sim 3\times 10^{-4}$ at $t_d = 9.8$ ps (see Supplementary Information on the linearity of the frequency conversion process). The theoretically calculated magnitude at $\omega_m$ based on Eq. 2 is plotted in Fig. 5e for a comparison. Theoretical and experimental magnitude plots all show a fringe pattern with a dominant evolution frequency at $2\omega_{1i}$, which results from the interference between the $\omega_m - \omega_i$ and $\omega_m + \omega_i$ components.

Measured and simulated data for the case of superposition between up- and down-conversion are shown in Figs. 5f-j. Here, a dual-band multi-cycle pulse whose centre wavelengths are $\omega_{i1}$ and $\omega_{i2}$ was used to excite the original modes. Figure 5f shows the complex amplitude at $\omega_m$ and its projected trajectories, which exhibit a complicated pattern similar to that shown in Fig. 4d. Again, the similarity between the measured and calculated data can be confirmed in the evolutionary spectrum (Fig. 5g); in this case, the Fourier-transformed spectrum with respect to the time delay exhibits four frequency components: $\omega_m \pm \omega_{i1}$ and $\omega_m \pm \omega_{i2}$. The magnitude spectra and their values measured and calculated at $\omega_m$ are plotted as a function of the time delay in Figs. 5h-j. The power conversion efficiency, given in this case as $|E(\omega_m)|^2/(|E(\omega_{i1})|^2 + |E(\omega_{i2})|^2)$, is as high as $\sim 6 \times 10^{-4}$ at $t_d = 5.2$ ps. Theoretical and experimental magnitude plots show more complicated fringe patterns (relative to the single-band excitation case) because they originate from the interference between the four components $\omega_m \pm \omega_{i1}$ and $\omega_m \pm \omega_{i2}$. In particular, an oscillating component at a frequency of $\omega_{i2} - \omega_{i1}$ in the magnitude plot is the most prominent one because the $\omega_m - \omega_{i1}$ and $\omega_m - \omega_{i2}$ components are dominant terms in the complex amplitude of the converted wave.

Since our proposed conversion scheme does not rely on the nonlinearity of the constituent materials, the conversion efficiency is invariant with respect to the intensity of an input wave. Therefore, this method is expected to be particularly beneficial for the frequency conversion of waves with weak intensities, which is distinct from the case of frequency conversion in nonlinear materials[27] or nonlinear metamaterials[28-30]. Furthermore, the frequency of a converted wave and its efficiency is tailorable to a large degree because the conversion process does not require energy conservation between participating waves and because the metasurface can be reconfigured to exhibit temporally distinct dispersive responses. This type of engineering degree of freedom was not achievable in previous plasma-based apparatus or photonic crystal devices. Last but not least, the phase of the converted wave can be fully controlled by $2\pi$ radians, and this full phase control is expected to enable new phase-sensitive applications. Although the current demonstration was limited to the case of a step-like temporal variation, further sophistication could be achieved with other types of temporal variations, such as Dirac delta or sinusoidal functions. Specifically, with Dirac-delta-like temporal variation, along with pulse-shaping technology, an artificial coherent control system might be realized with metasurfaces.

**Methods**

**Measurement of frequency conversion in time-variant metasurfaces.** All measurements described in the main manuscript were obtained using a typical THz-TDS system based on a Ti:sapphire regenerative amplifier operating at a repetition rate of 250 kHz. A parabolic mirror with a hole was used to bypass the pump pulse while focusing the THz pulse onto the metasurface. The optical pump pulse and the THz input pulse were collinearly delivered to the metasurface with this arrangement. The pulse width and the centre wavelength of the pump pulses were 50 fs and 800 nm, respectively. The spot diameter of the pump pulse on the metasurface was approximately 4 mm, which was sufficiently large to cover the region of the THz pulse excitation. To measure the transmission spectra, such as those as shown in Fig. 3, we used a commercially available large-area photoconductive antenna as a THz emitter. To prepare the multi-cycle pulses used to produce the data shown in Fig. 5, we chose a lithium niobate crystal as a THz emitter due to its broad spectral bandwidth. Waveforms of these THz input pulses are shown in Supplementary Information. For the detection of the THz waves, an electro-optic sampling technique with a 2-mm-thick <110>-oriented ZnTe crystal was used.

**Metasurface fabrication.** The metasurface was fabricated on a 625-μm-thick semi-insulating, undoped GaAs substrate. An array of 200-nm-thick gold SRRs was patterned on the substrate using a lift-off process. Next, a 300-nm-thick $SiO_2$ layer was deposited for passivation. On top of the passivation layer, 10 pairs of 94-nm-thick $TiO_2$ and 136-nm-thick $SiO_2$ layers were subsequently deposited. Electron-beam evaporation was used to deposit the $SiO_2$ and $TiO_2$. To create an opening on these alternating dielectric layers, reactive ion etching (RIE) with tetrafluoromethane ($CF_4$) was used. All patterns in the process were defined with negative photoresist (AZ® nLOf 2035, MicroChemicals GmbH, Ulm, Germany).

**Acknowledgements**

We thank Dr. Sunkyu Yu and Dr. Xianji Piao for helpful discussions. This work was supported by the Basic Science Research Program (2012R1A2A1A03670391), the Nano Material Technology Development Program (2014039957), the Pioneer Research Center Program (2014M3C1A3052537) and the Quantum Metamaterials Research Center Program (No. 2015001948) through a National Research Foundation of Korea (NRF) grant funded by the Korean government (MSIP). This work was also supported by the Center for Advanced Meta-Materials (CAMM) funded by the Ministry of Science, ICT and Future Planning as a Global Frontier Project (CAMM-2014M3A6B3063709).

**Author contributions**

K.L., J.S. and B.M. conceived the original idea. J.S., B.K. and J.P. fabricated metasurface samples and THz band-pass filters. K.L., J.S. and J.P. performed the measurements. J.S. performed the FDTD simulations. K.L., J.S., J.P., W.J., F.R. and B.M. discussed the theoretical and experimental results. K.L., J.S. and B.M. wrote the manuscript, and all authors provided feedback.

**Competing financial interests**

The authors declare no competing financial interests.

**Figure captions**

**Figure 1 The principle of frequency conversion in time-variant metasurfaces.** Conceptual schemes, timing diagrams, and corresponding amplitude transmissions are shown for three regimes of operation. (a) For the regime in which the temporal boundary is far behind of the input pulse, the amplitude transmission shows two dips corresponding to the original modes of the two SRRs. (b) For the regime in which the temporal boundary disturbs the original modes driven by the input pulse, a new frequency component is radiated from an oscillating merged mode, and the transmission at the merged frequency becomes larger than unity. (c) For the regime in which the temporal boundary is ahead of the input pulse, the amplitude transmission shows a single dip corresponding to the merged mode. Amplitude transmissions in the bottom panels were calculated from FDTD simulations.

**Figure 2 Structures and steady-state responses of a proposed time-variant metasurface.** (a) Schematic illustration of the metasurface structure. Photo-carrier injection occurs only in the dark grey region as the other area is screened from pump pulses by the patterned $SiO_2/TiO_2$ multilayer. (b) Optical micrograph of the metasurface taken before the deposition of the $SiO_2/TiO_2$ multilayer. (c) Experimentally measured amplitude transmission through the metasurface for various pump fluence levels. (d) Numerically calculated amplitude transmission of the metasurface using FDTD simulations.

**Figure 3 Time-delay-dependent amplitude transmission measured at a pump fluence of 10 μJ/cm².** (a) Amplitude transmission spectra of the metasurface as a function of time delay. (b) Amplitude transmission spectra for three distinct time delays at $t_d$ = -4 ps (black), $t_d$ = 2 ps (red), and $t_d$ = 15 ps (blue). (c) Amplitude transmission at $\omega_m$ (0.92 THz) as a function of time delay. The inset shows the expanded view between $t_d$ = 1 ps and $t_d$ = 7 ps. The dotted line is the time-delay-dependent amplitude transmission calculated using Eq. 2.

**Figure 4 Trajectories of complex amplitudes for the converted field with respect to the time delay.** (a) Trajectory of the complex amplitude for the converted field from the lower frequency to the merged frequency. The two counter-clockwise rotating components at frequencies of $\omega_m - \omega_i$ and $\omega_m + \omega_i$ are superposed to produce a phasor trajectory evolving similarly to the shape of a flower. (b) Trajectory of the complex amplitude for the converted field from the higher frequency to the merged frequency. The trajectory evolves similarly to the shape of a starfish due to oppositely rotating components at frequencies of $\omega_m - \omega_i$ and $\omega_m + \omega_i$. (c) Trajectory of the complex amplitude for the converted field from the lower and higher frequencies to the merged frequency. The complex amplitude is the superposition of (a) and (b). (d) Trajectory of the complex amplitude for the converted field with realistic values of $\alpha$ and $\phi$.

**Figure 5 Complex amplitude of the converted field and its magnitude for multi-cycle input pulses of a single band (a-e, with a centre frequency of $\omega_{i1}$) and dual bands (f-j, with centre frequencies of $\omega_{i1}$ and $\omega_{i2}$).** (a) Complex amplitude of the converted field at $\omega_m$ plotted as a function of the time delay (blue) and corresponding projections on each of the planes. The inset shows the spectrum of the input pulse. (b) Evolutionary spectrum of the complex amplitude. A Fourier transform is performed with respect to the time delay. (c) Magnitude of the complex amplitude for a spectral range of 0.8-1THz. (d) Magnitude of the complex amplitude at $\omega_m$. (e) Theoretical calculation of the magnitude at $\omega_m$. (f) Complex amplitude of the converted field at $\omega_m$ plotted as a function of the time delay (blue) and corresponding projections on each of the planes. The inset shows the spectrum of the input pulse. (g) Evolutionary spectrum of the complex amplitude. (h) Magnitude of the complex

amplitude for a spectral range of 0.8-1THz. (i) Magnitude of the complex amplitude at $\omega_m$. (j) Theoretical calculation of the magnitude at $\omega_m$.

# Figures

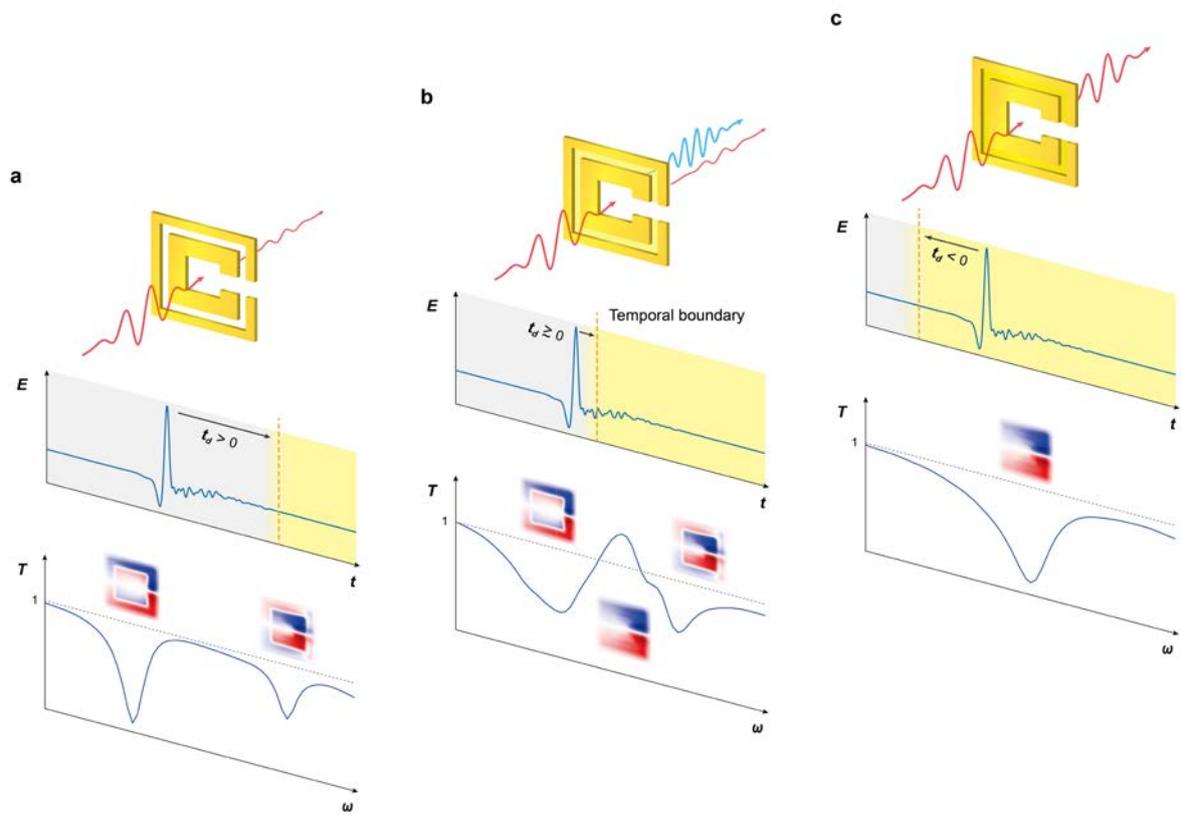

Fig. 1 Lee et al.

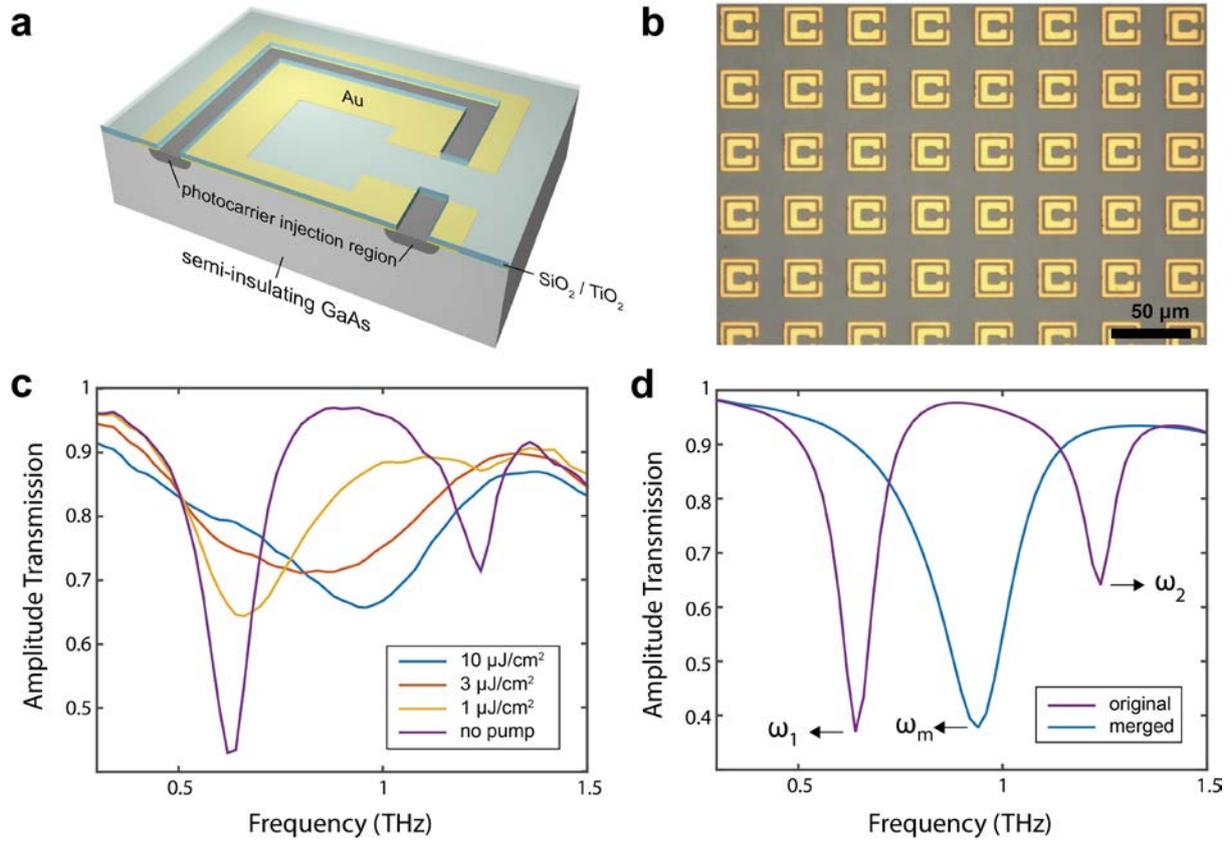

Fig. 2 Lee et al.

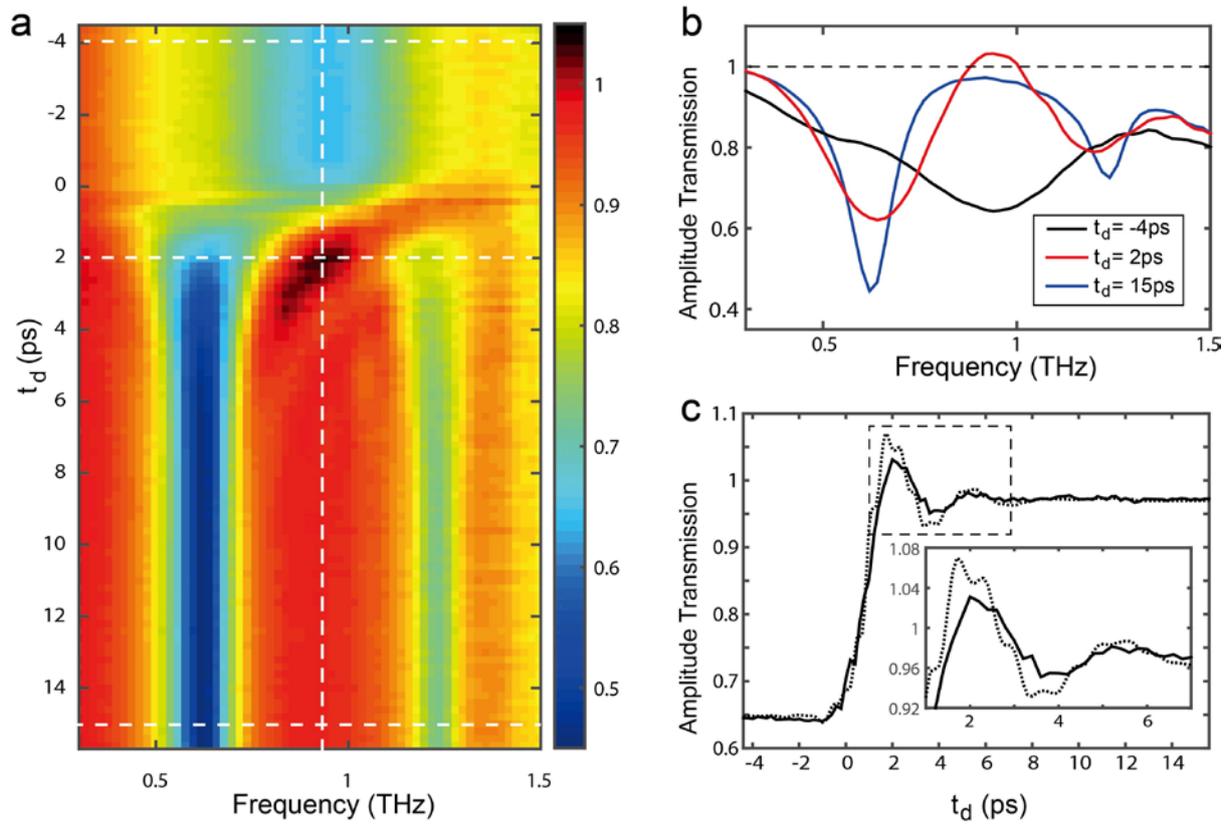

Fig. 3 Lee et al.

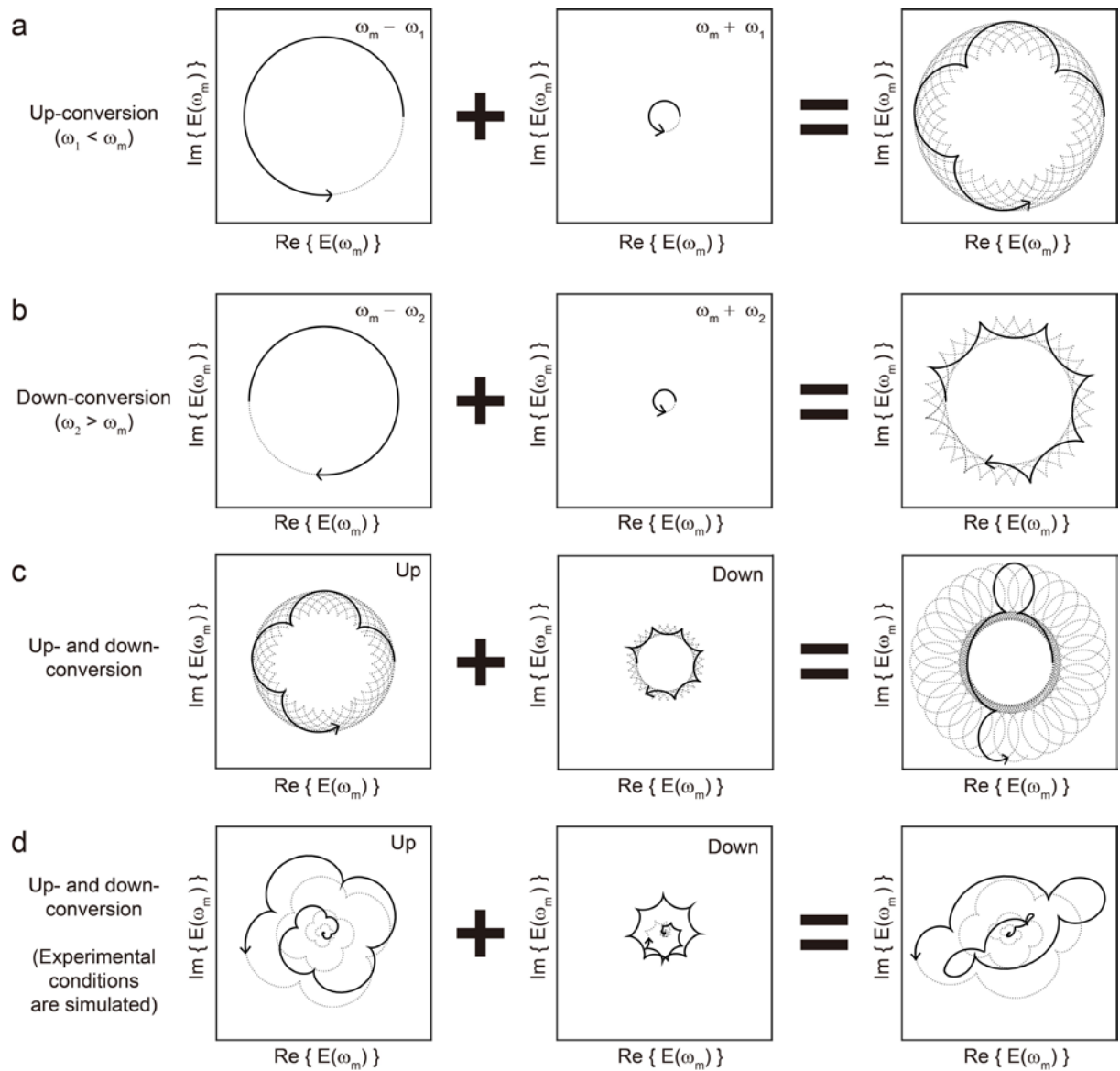

Fig. 4 Lee et al.

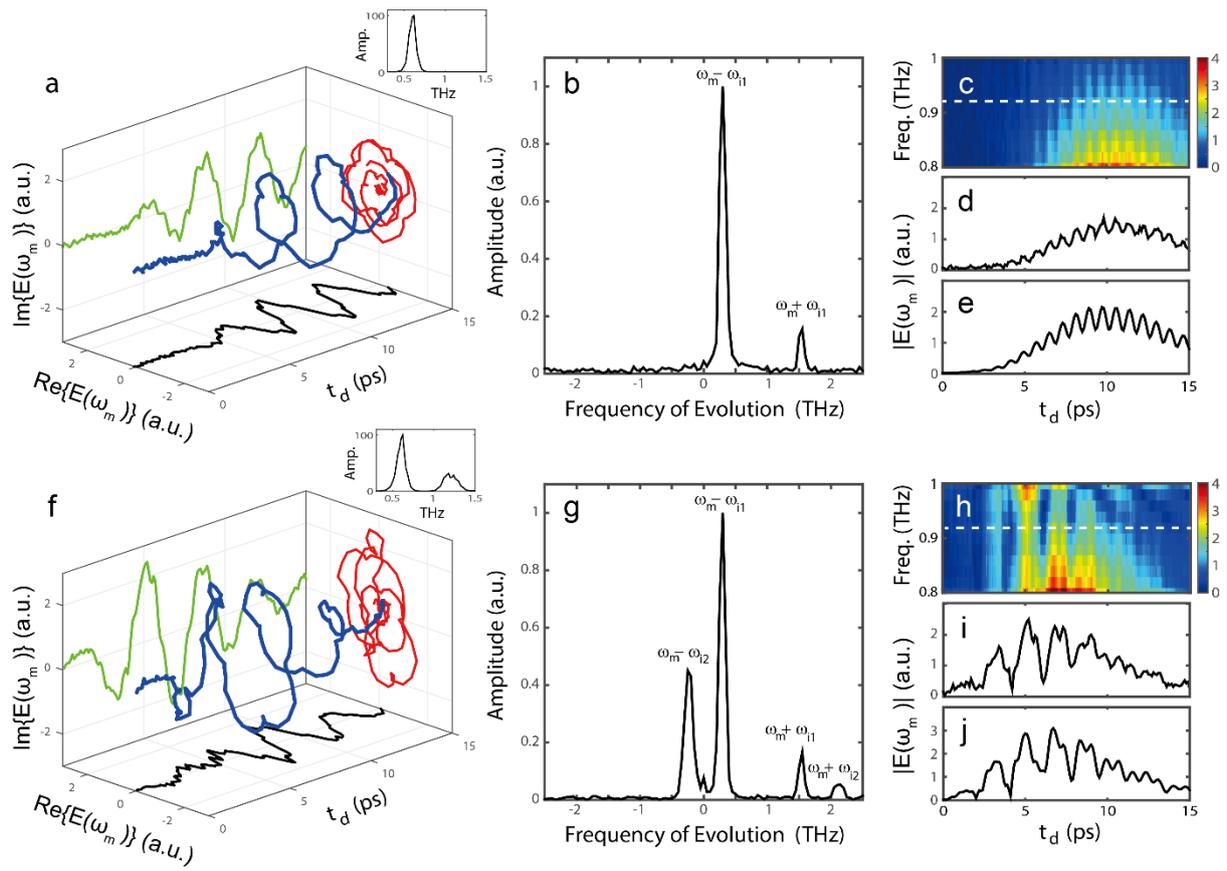

Fig. 5 Lee et al.

# Supplementary Information for "Linear frequency conversion via sudden merging of resonances in time-variant metasurfaces"

Kanghee Lee[1,†], Jaehyeon Son[1,†], Jagang Park[1], Byungsoo Kang[1], Wonju Jeon[1], Fabian Rotermund[2] & Bumki Min[1,*]

[1]Department of Mechanical Engineering, Korea Advanced Institute of Science and Technology (KAIST), Daejeon 305-751, Republic of Korea

[2]Department of Physics, Korea Advanced Institute of Science and Technology (KAIST), Daejeon 305-751, Republic of Korea

In this supplementary information, we present details of theoretical analysis and numerical simulations, metasurface fabrication, and additional measurement for "Linear frequency conversion via sudden merging of resonances in time-variant metasurfaces".

# 1. Theoretical derivation of formulae related to frequency conversion

This section is devoted to a detailed theory on frequency conversion briefly introduced in the main text. To simplify the problem, several assumptions were made. First, the metasurface was approximated as an infinitesimally thin, homogeneous but dispersive surface that is sandwiched between air and a non-magnetic dielectric substrate with a refractive index of $n$. Second, a linearly polarized plane wave was assumed to be incident normally on the metasurface. Under these conditions, incident ($\tilde{E}_i(\omega), \tilde{H}_i(\omega)$), reflected ($\tilde{E}_r(\omega), \tilde{H}_r(\omega)$), and transmitted ($\tilde{E}_t(\omega), \tilde{H}_t(\omega)$) fields can be defined near the metasurface as depicted in Fig. S1. From boundary conditions of Maxwell's equations, the relationship between the fields can be written as follows:

$$\tilde{E}_i(\omega) + \tilde{E}_r(\omega) - \tilde{E}_t(\omega) = 0 \tag{1}$$

$$\tilde{H}_i(\omega) - \tilde{H}_r(\omega) - \tilde{H}_t(\omega) = \tilde{J}(\omega) \tag{2}$$

where $\tilde{J}$ is the surface current density of the metasurface. For this simple case, the magnetic fields are found to be proportional to the corresponding electric fields with a scaling factor:

$$\tilde{H}_{i,r}(\omega) = \tilde{E}_{i,r}(\omega)/Z_0 \tag{3}$$

$$\tilde{H}_t(\omega) = n\tilde{E}_t(\omega)/Z_0 \tag{4}$$

where $Z_0$ is the impedance of free space. Equations (1) to (4) are combined to obtain the electric field of the transmitted wave, $\tilde{E}_t(\omega)$, as follows:

$$\tilde{E}_t(\omega) = \frac{2}{1+n}\left[\tilde{E}_i(\omega) - \frac{Z_0}{2}\tilde{J}(\omega)\right]. \tag{5}$$

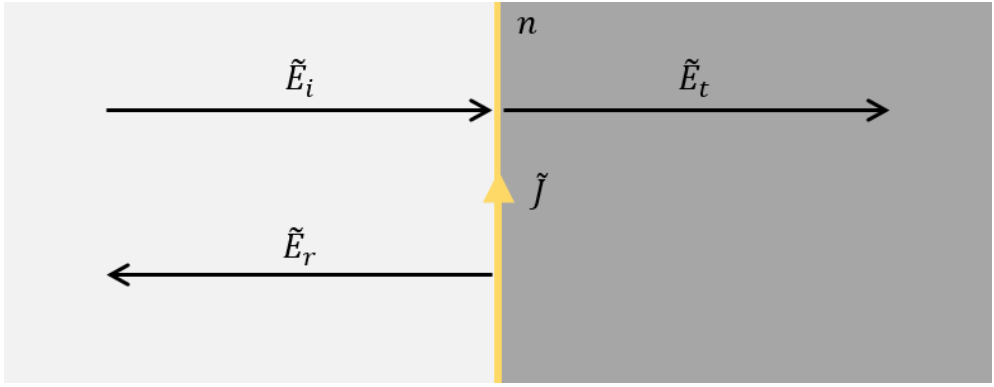

**Figure S1. Simplified model of a metasurface for the frequency conversion analysis.**

This Eq. (5) is Eq. (1) in the main text. Before analysing time-varying behaviours, let us investigate the steady-state responses in further detail. The surface current density can be rewritten in terms of the transmitted wave as follow:

$$\tilde{J}(\omega) = \tilde{\sigma}(\omega)\tilde{E}_t(\omega) \tag{6}$$

Combining Eqs. (5) and (6) gives the relationship between the surface current density and the electric field of the incident wave:

$$\tilde{J}(\omega) = \frac{2\tilde{\sigma}(\omega)}{1 + n + Z_0\tilde{\sigma}(\omega)} \tilde{E}_i(\omega) \tag{7}$$

Here, the metasurface has two steady states with different surface conductivities. One is called the original state and has a surface conductivity of $\tilde{\sigma}_o(\omega)$; the other is called the merged state and has a surface conductivity of $\tilde{\sigma}_m(\omega)$. From Eq. (7), the linear response function of each steady state can be defined as

$$\theta_{o,m}(t) = \mathcal{F}^{-1}\left\{\frac{2\tilde{\sigma}_{o,m}(\omega)}{1 + n + Z_0\tilde{\sigma}_{o,m}(\omega)}\right\} \tag{8}$$

where $\mathcal{F}^{-1}$ denotes the inverse Fourier transform. The surface current density in the time domain can be expressed in the form of convolution integral as follows:

$$J_{o,m}(t;E) = \int_{-\infty}^{\infty} \theta_{o,m}(t-\tau)E(\tau)d\tau \qquad (9)$$

where $J_o(t;E)$ and $J_m(t;E)$ are the surface current densities of the original and merged states, respectively, for the arbitrary incident wave with electric field $E(t)$. The causality imposes the condition $\theta_{o,m}(t)|_{t<0} = 0$. It is useful to define a *virtual* input field, $E_v(t)$ that satisfies

$$J_o(t;E_i) = J_m(t;E_v) \quad \text{for all } t. \qquad (10)$$

In other words, the merged state response with the virtual input field gives the same surface current density as the original state response with the actual input field, $E_i(t)$.

Now, let us consider the time-varying situation, in which the state of the metasurface *abruptly* switches from the original to the merged state at $t = t_d$. For this case, the *dynamic* surface current density formula applicable over the entire time domain can be derived from *steady-state* responses corresponding to two distinct time periods ($t < t_d$: original state and $t > t_d$: merged state) as follows:

$$J(t) = [1 - u(t - t_d)] J_o(t;E_1) + u(t - t_d) J_m(t;E_2) \qquad (11)$$

where $u(t)$ is the unit step function, $E_1(t) = E_i(t)$ for all $t$, and $E_2(t) = E_i(t)$ for $t > t_d$. To determine $E_2(t)$ for $t < t_d$, continuous conditions on physically observable parameters at the temporal boundary must be imposed. The following relation is a sufficient condition to satisfy the continuity conditions on *all* physically observable parameters:

$$J_o(t;E_1) = J_m(t;E_2) \quad \text{for } t < t_d \qquad (12)$$

Since $E_1(t) = E_i(t)$, Eq. (12) implies that $E_2(t) = E_v(t)$ for $t < t_d$; thus,

$$E_2(t) = [1 - u(t - t_d)]E_v(t) + u(t - t_d)E_i(t) \tag{13}$$

By combining Eqs. (11) to (13), we have

$$J(t) = J_m(t; E_2) = \int_{-\infty}^{\infty} \theta_m(t - \tau)[E_v(\tau) + u(\tau - t_d)(E_i(\tau) - E_v(\tau))]d\tau \tag{14}$$

Performing the Fourier transform on Eq. (14) and using the definition in Eq. (10) yields Eq. (15):

$$\tilde{J}(\omega) = \tilde{\theta}_o(\omega)\tilde{E}_i(\omega) + \tilde{\theta}_m(\omega) \int_{t_d}^{\infty} [E_i(t) - E_v(t)]e^{-i\omega t}dt, \tag{15}$$

which is Eq. (2) in the main text.

Let us further consider the situation in which a narrow-band pulse is incident on the metasurface as in the experiments described in the main text. In this case, the electric field of the incident wave can be approximated as follows:

$$E_i(t) \approx \alpha(t) \cos \omega_i t \tag{16}$$

where $\omega_i$ is the carrier frequency in radians. A slowly varying envelope is described by $\alpha(t)$, and this envelope is assumed to approach to zero as time approaches infinity or negative infinity, i.e., $\lim_{t \to \pm\infty} \alpha(t) = 0$. From Eq. (10), the difference between the input and virtual input field can be approximated as,

$$E_i(t) - E_v(t) \approx |\beta|\alpha(t)\cos(\omega_i t + \phi) \tag{17}$$

where $\beta = 1 - \frac{\tilde{\theta}_o(\omega_i)}{\tilde{\theta}_m(\omega_i)}$ and $\phi$ is the phase angle of $\beta$. Then, the integration in Eq. (18) can be approximated as,

$$\int_{t_d}^{\infty} [E_i(t) - E_v(t)] e^{-i\omega t} dt \approx \int_{t_d}^{\infty} |\beta| \alpha(t) \cos(\omega_i t + \phi) e^{-i\omega t} dt$$

$$= |\beta| \left[ -\frac{\alpha(t)}{2i} \left( \frac{e^{-i\phi - i(\omega+\omega_i)t}}{\omega + \omega_i} + \frac{e^{i\phi - i(\omega-\omega_i)t}}{\omega - \omega_i} \right) \right]_{t_d}^{\infty}$$

$$+ |\beta| \int_{t_0}^{\infty} \frac{d\alpha}{dt} \frac{1}{2i} \left[ \frac{e^{-i\phi - i(\omega+\omega_i)t}}{\omega + \omega_i} + \frac{e^{i\phi - i(\omega-\omega_i)t}}{\omega - \omega_i} \right] dt \qquad (18)$$

$$\approx |\beta| \alpha(t_d) \frac{1}{2i} \left[ \frac{e^{-i\phi - i(\omega+\omega_i)t_d}}{\omega + \omega_i} + \frac{e^{i\phi - i(\omega-\omega_i)t_d}}{\omega - \omega_i} \right].$$

We note that this approximation is valid when $|\omega_i - \omega|$ is not approaching zero. At frequencies far from the carrier frequency, the electric field of the incident wave becomes zero, i.e. $\tilde{E}_i(\omega) = 0$ and the electric field of the transmitted wave can be approximated as follows:

$$\tilde{E}_t(\omega) \approx -\frac{Z_0}{1+n} \tilde{\theta}_m(\omega) |\beta| \alpha(t_d) \frac{1}{2i} \left[ \frac{e^{-i\phi - i(\omega+\omega_i)t_d}}{\omega + \omega_i} + \frac{e^{i\phi - i(\omega-\omega_i)t_d}}{\omega - \omega_i} \right]. \qquad (19)$$

Equation (19) is Eq. (3) of the main text and is schematically described in Fig. 4 of the main text. According to Eq. (19), the magnitude of the converted field has a scaling factor, $|\beta|$, which is a function of the carrier frequency of the narrow band input wave. In Fig. S2, $|\beta|$ is plotted with the amplitude transmission of the original state and merged state. In the plot, $|\beta|$ has two maximum points near the original resonance frequencies of $\omega_{1,2}$, but they are not located exactly at the original resonance frequencies. The shifts are originated from the phase difference of $\tilde{\theta}_o(\omega_i)$ and $\tilde{\theta}_m(\omega_i)$.

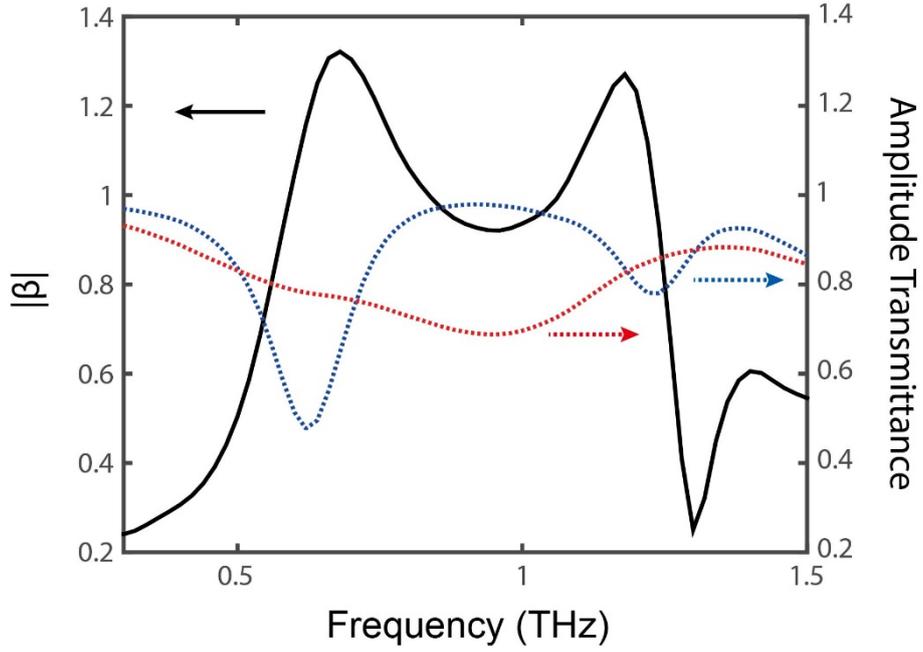

**Figure S2.** The scaling factor, $|\boldsymbol{\beta}|$, calculated from the measured data is plotted in the black solid line. The blue dotted line and the red dotted line indicate transmission of the original state and merged state, respectively.

## 2. Definition and identification of the zero delay time, $t_d = 0$

The zero delay time, $t_d = 0$, between an incident THz pulse and an optical pumping pulse was assigned experimentally as follows. First, a peak electric field of the transmitted THz pulse *through a bare GaAs wafer* was measured as a function of the time delay. Figure S3 shows such a peak electric field, $E_{\text{peak}}$, as a function of $t_d$. From this plot, the zero delay time, $t_d = 0$, is defined as a delay time point when $E_{\text{peak}} = (\max\{E_{\text{peak}}\} + \min\{E_{\text{peak}}\})/2$. Here, $\max\{E_{\text{peak}}\}$ and $\min\{E_{\text{peak}}\}$ denote the (averaged) maximum and minimum values of $E_{\max}$, respectively, for the measured range of delay times. For the cases with multi-cycle THz input pulses, the zero delay time was similarly defined but without including band-pass filters in the pathway. The inclusion of filters delayed arrival of the THz pulse so that the actual overlapping range of the input pulses and the time boundary was located in $t_d > 0$ ps for these cases.

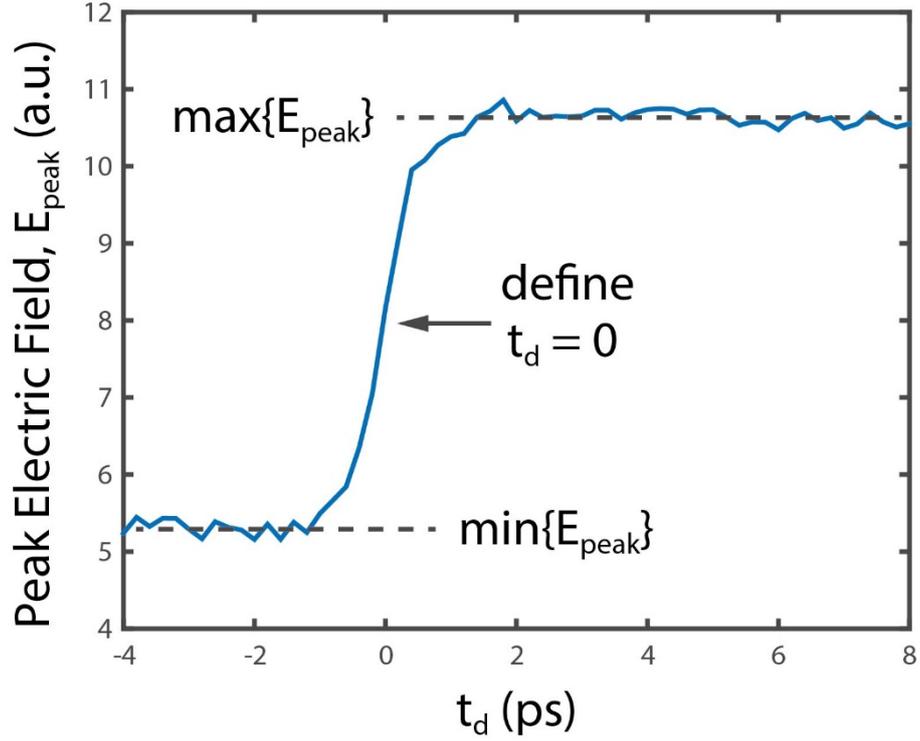

**Figure S3. The peak electric field, $S$, of the transmitted THz pulse as a function of $t_d$.** The zero delay time, $t_d = 0$, was defined at a time delay point satisfying $E_{\text{peak}} = (\max\{E_{\text{peak}}\} + \min\{E_{\text{peak}}\})/2$.

## 3. Confirmation of linearity in frequency conversion

To check the linearity of frequency conversion, additional measurements were conducted with dual-spectral-band multi-cycle THz input pulses. The relation between the powers of incident and converted THz waves is plotted in Fig. S4. Here, the input power is calculated as the sum of the power spectral density of the incident THz waves, while the converted THz power refers to the sum of the power spectral density of the transmitted THz waves near the merged frequency from 0.925 to 0.975 THz. Both the time delay and the optical pumping fluence (10 µJ/cm$^2$) were fixed in this measurement. Figure S4 clearly shows that the converted THz power is linearly proportional to the incident THz power.

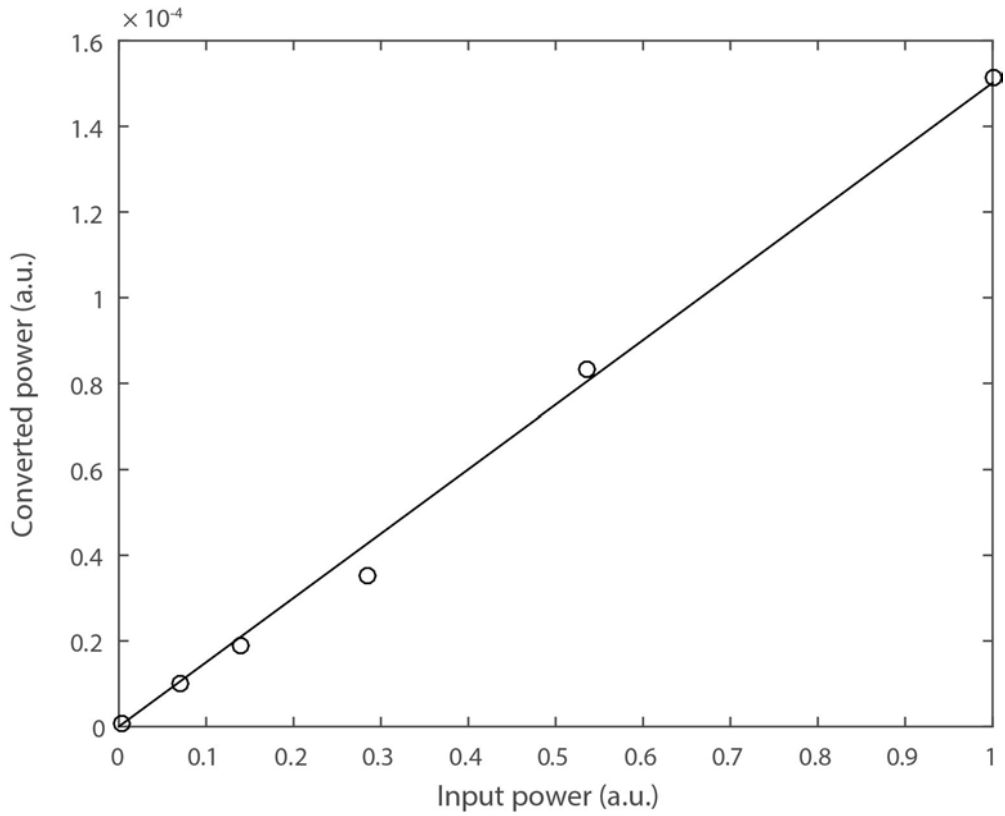

**Figure S4. Converted power versus input power.** The circles represent measured data, and the solid line is the corresponding linear fit.

## 4. Geometry of the meta-atoms and fabrication

Figure S5a shows the geometries and relevant parameters of SRRs in a unit cell with a lattice constant of 40 μm. Two SRRs are designed to have distinct side lengths of $L_1$ = 24 μm and $L_2$ = 16 μm, widths of $w_1$ = 2 μm and $w_2$ = 4 μm, and the same split length of $c$ = 4 μm. The gap distance ($g$) between the two SRRs is set to 2 μm. Figure S5b shows the geometry and relevant parameters of the opening in multiple dielectric layers (composed of alternating $TiO_2$ and $SiO_2$ layers). The opening was formed by etching the dielectric layers in the shape of an SRR (with $L_o$ = 22 μm, $w_o$ = 4 μm and $c_o$ = 6 μm). Notably, $w_o$ and $c_o$ are wider than $g$ and $c$, respectively, to minimize unwanted side effects, such as carrier diffusion, misalignment during fabrication, and diffraction of the optical pump beam. The sectional dissected view of the metasurface is also shown in Fig. S5c. The metasurface was fabricated on a semi-insulating GaAs wafer.

Using conventional photolithography and the lift-off process, 200-nm-thick gold was patterned to form an array of double SRRs on the substrate. To passivate the SRRs, 300-nm-thick $SiO_2$ was deposited after the lift-off process. Alternatively, ten pairs of $TiO_2$ and $SiO_2$ films were deposited (each with 94 and 136 nm thick) on the $SiO_2$ passivation layer, forming a Bragg mirror to block over 99% of the incident light at the 800 nm wavelength. RIE with $CF_4$ was used to form the opening, i.e., the pump-through window, on the dielectric layers.

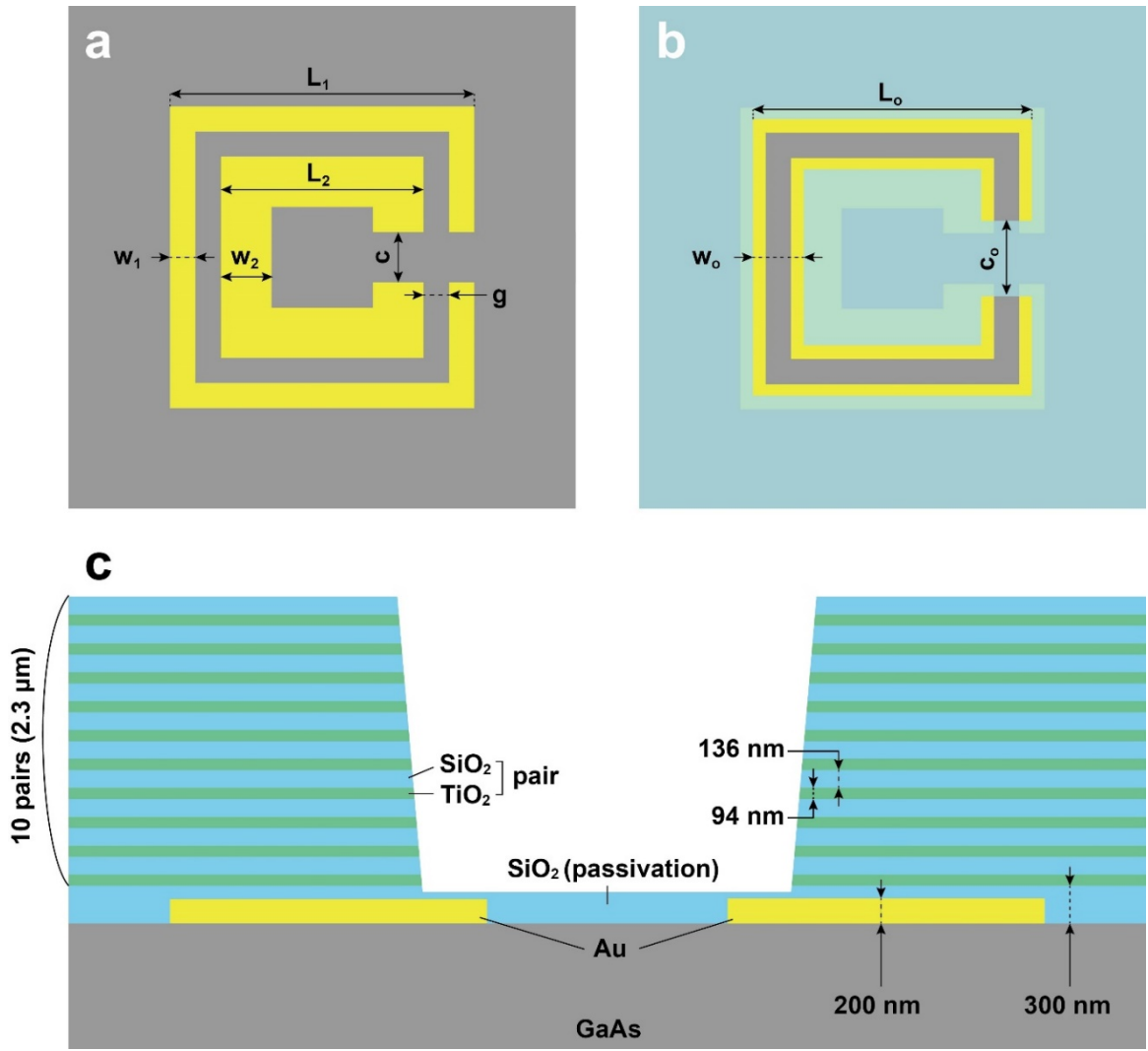

**Figure S5. Schematics of the fabricated metasurface**. (a) Top view of the two SRRs in a unit cell: $L_1 = 24$ μm, $L_2 = 16$ μm, $w_1 = 2$ μm, $w_2 = 4$ μm, $c = 4$ μm, and $g = 2$ μm. (b) Top view of the opening on the dielectric layer: $L_o = 22$ μm, $w_o = 4$ μm, and $c_o = 6$ μm. c. Schematic sectional view of the metasurface.

## 5. Characteristics of an input THz pulse without filtering

The time trace of the electric field of an input THz pulse and the corresponding spectrum (inset) are shown in Fig. S6. This broadband sub-cycle input pulse was used to produce the data shown in Fig. 3 of the main text.

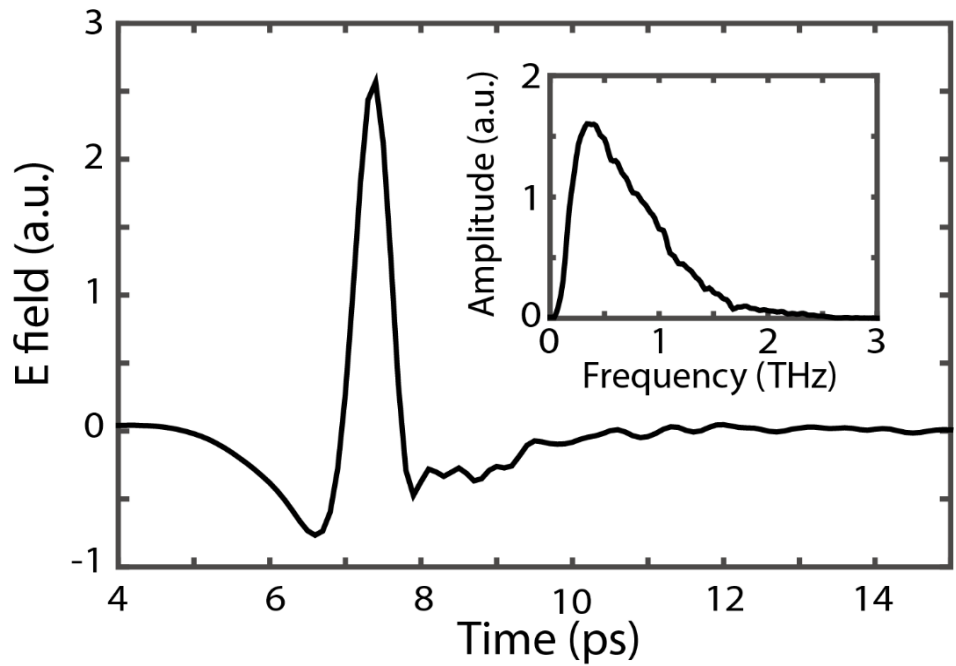

**Figure S6. Time trace of the THz input pulse.** A broadband sub-cycle input pulse was used to measure the amplitude transmission shown in Fig. 3 of the main text. The inset shows the pulse spectrum.

## 6. Characteristics of THz band-pass filters

To produce a multi-cycle THz input pulse for the measurement of data shown in Fig. 5 of the main text, a broadband sub-cycle input pulse was transmitted through a series of band-pass filters. Geometries and characteristics of the designed band-pass filters are shown in Fig. S7. Band-pass filters with a single passband at $\omega_1$ and those with dual passbands at $\omega_1$ and $\omega_2$ were designed with complementary cross-type resonators. Filters were fabricated using conventional photolithography and the lift-off process; an array of complementary crosses with a width ($w_1$) of 31 μm, a length ($L_1$) of 220 μm, and a lattice constant ($a$) of 250 μm was patterned on 2-µm-thick polyimide for the band-pass filter with a single passband at $\omega_1$, while arrays of two distinct crosses (widths of $w_{1,2}$ = 24 and 31 µm, lengths of $L_{1,2}$ = 220 and 103 µm, and a lattice constant $a$ = 250 µm) were patterned for the band-pass filter with dual passbands at $\omega_1$ and $\omega_2$. Amplitude transmissions through the single- and dual-passband filters are plotted in Figs. S7b and S7d, respectively. To eliminate spectral components near the merged frequency at $\omega_m$ from the input THz pulse, several sets of identical filters were stacked to enhance spectral selectivity. For the results shown in Figs. 5a-e of the main text for a single-spectral-band multi-cycle input pulse, four filters with a single passband at $\omega_1$ and two filters with dual passbands at $\omega_1$ and $\omega_2$ were stacked with a spacing of ~ 6 mm between the constituent filters. The corresponding amplitude transmission through the stacked filters is shown in Fig. S7e. For the dual-band input pulse in Figs. 5f-j of the main text, four filters with dual passbands at $\omega_1$ and $\omega_2$ were stacked, and the amplitude transmission is shown in Fig. S7f.

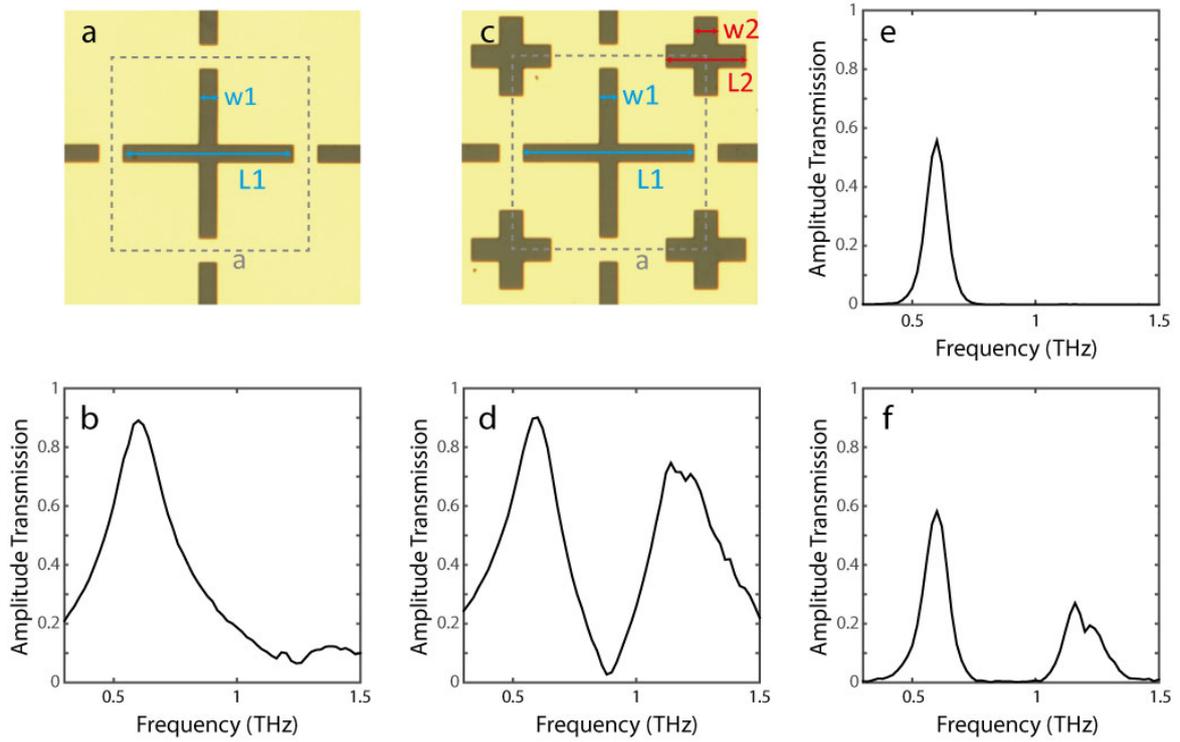

**Figure S7. Characteristics of THz band-pass filters used for the generation of multi-cycle input pulses.** (a) Optical micrograph of the band-pass filter with a single passband at $\omega_1$. (b) Amplitude transmission through the band-pass filter with a single passband at $\omega_1$. (c) Optical micrograph of the band-pass filter with dual passbands at $\omega_1$ and $\omega_2$. (d) Amplitude transmission through the band-pass filter with dual passbands at $\omega_1$ and $\omega_2$. (e) Amplitude transmission through the filter configured to produce a single-spectral-band multi-cycle input pulse described in the main text. Specifically, four filters with a single passband at $\omega_1$ and two filters with dual passbands at $\omega_1$ and $\omega_2$ were stacked to implement the composite filter. (f) Amplitude transmission through the filter was configured to produce a dual-spectral-band multi-cycle input pulse described in the main text. In this case, four filters with dual passbands at $\omega_1$ and $\omega_2$ were stacked to create the composite filter.

## 7. Characteristics of multi-cycle THz pulses

The electric field time trace of a single-spectral-band multi-cycle input pulse is plotted in Fig. S8a. This THz input pulse was used in the actual measurement and theoretical calculations as shown in Figs. 5a-e of the main text. Figure S8b shows the amplitude spectrum of the input pulse, which is also shown in the inset of Fig. 5a of the main text. The time trace of the electric field of a dual-spectral-band multi-cycle input pulse is also plotted in Fig. S9a. This dual-spectral-band input pulse was used in the actual measurement and theoretical calculations as shown in Figs. 5f-j of the main text. Reconstructed time traces for $\omega_{i1}$(red) and $\omega_{i2}$(blue) spectral components, which can be obtained using the Fourier anaylsis, are shown in Fig. S9b. Envelops and phases of these reconstructed time traces were used to make plots in Fig. 4d of the main text. Figure S9c shows the amplitude spectrum of the dual spectral band input, which is also shown in the inset of Fig. 5e of the main text.

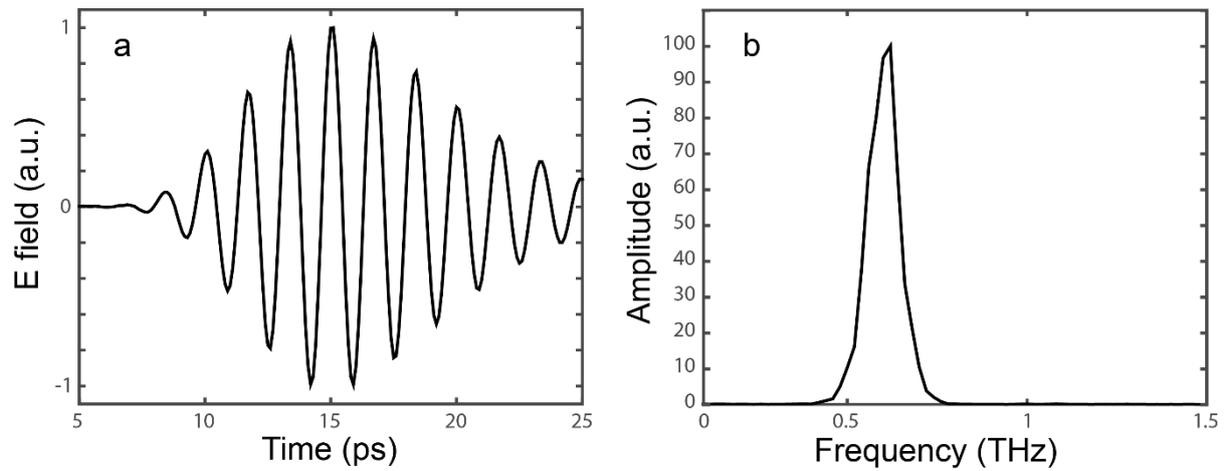

**Figure S8.** (a) Electric field time trace of a single-spectral-band multi-cycle THz input pulse. (b) The corresponding spectrum.

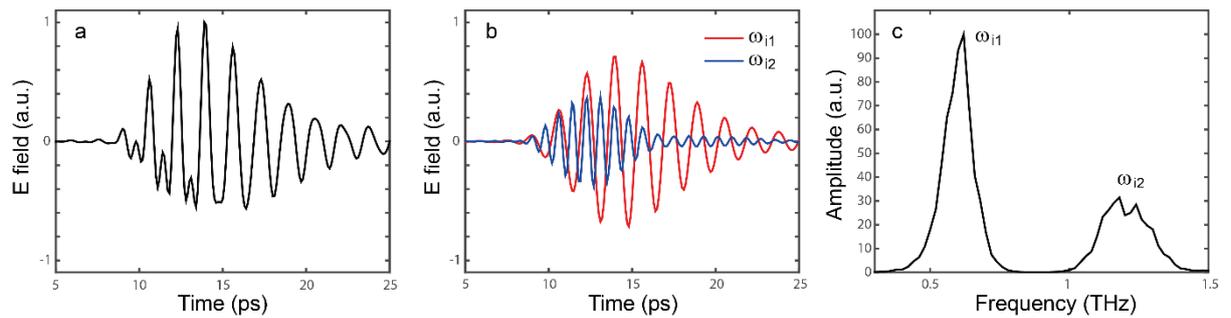

**Figure S9.** (a) Time traces of the electric field of a dual-spectral-band multi-cycle THz input pulse. (b) Reconstructed time traces for $\omega_{i1}$(red) and $\omega_{i2}$(blue) spectral components obtained using the Fourier analysis. (c) Amplitude spectrum of the dual-spectral-band multi-cycle input pulse.

## 8. FDTD simulation with time-varying materials and its verification

We used commercial software (Lumerical FDTD Solutions) and its built-in custom material system (flexible material plugins) to model a time-varying scenario in the simulation, which will be shown in the following section. To verify our method, we first reproduced the reflection and transmission of a pulse at a temporal boundary, for which analytical solutions were derived in a reference [1]. As shown in Fig. S10, our simulated data fit well with the analytically obtained solutions.

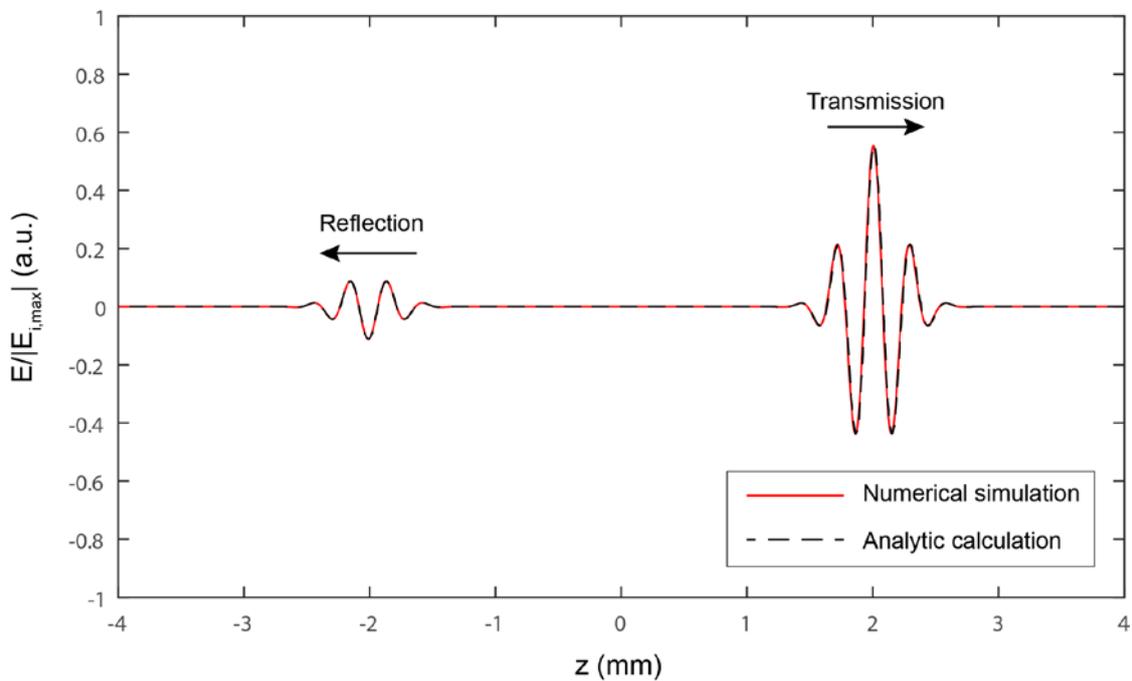

**Figure S10. Comparison between the time-varying FDTD simulation and the analytical calculation.** A wave packet is travelling in the positive $z$ direction inside a time-varying dielectric medium. The refractive index of the dielectric medium changes abruptly from 1 to 1.5 when the centre of the wave packet reaches $z = 0$. The plot shows the spatial profile of the electric fields of reflected and transmitted waves after 10 ps after interaction with the temporal boundary.

### 9. Reproduction of the measured data with time-varying FDTD simulations

The measured experimental data shown in Fig. 3 and Fig. 5 of the main text are reproduced here using previously introduced time-varying FDTD simulations. The characteristics of time-variant metasurfaces are reproduced in Figs. S11 and Fig. S12; excellent qualitative agreement to the data shown in Figs. 3 and Fig. 5 of the main text can be confirmed.

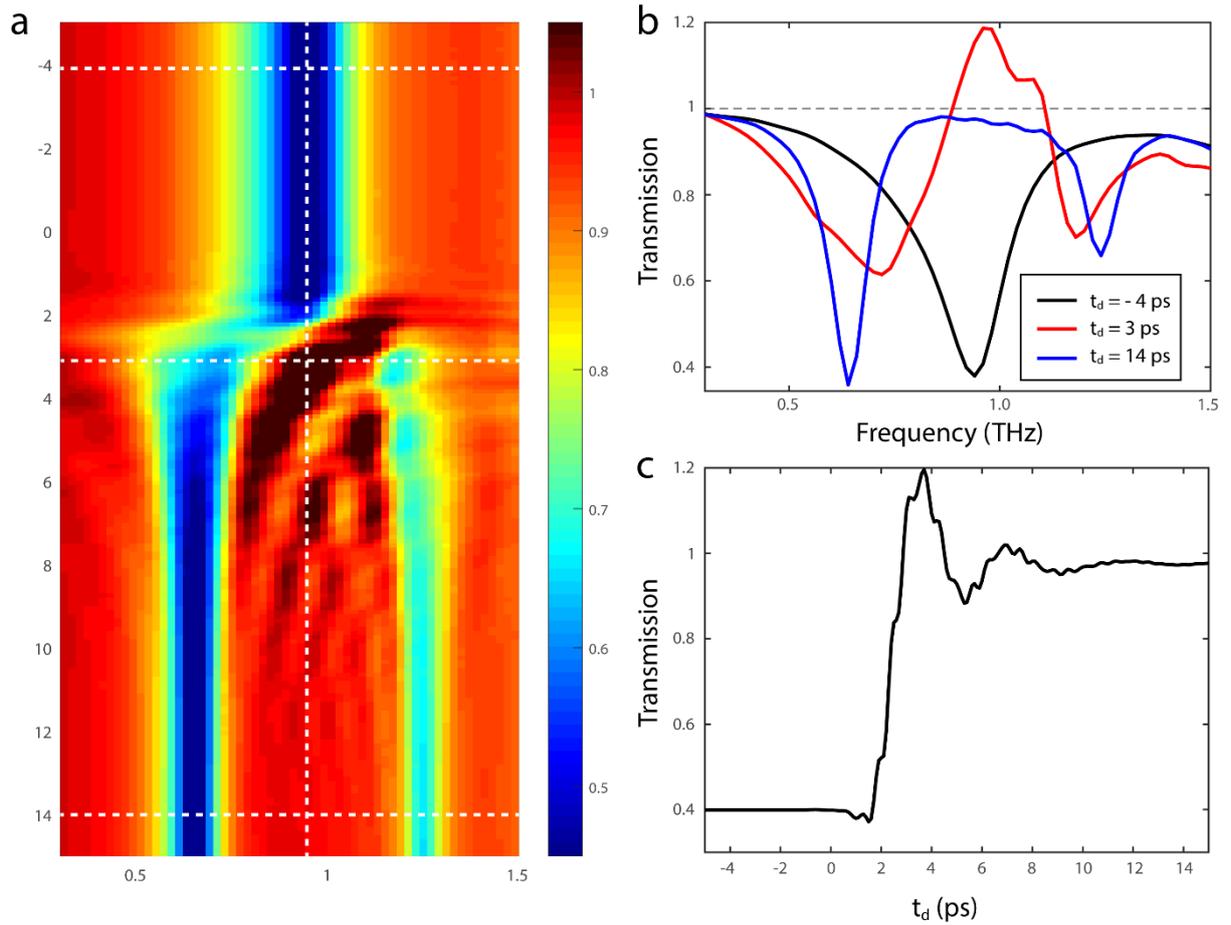

**Figure S11. Time-delay-dependent transmission in the time-varying FDTD simulation.** (a) Amplitude transmission spectra of the metasurface as a function of time delay. (b) Amplitude transmission spectra for three distinct time delays at $t_d =$ -4 ps (black), $t_d =$ 3 ps (red), and $t_d =$ 14 ps (blue). (c) Amplitude transmission at $\omega_m$ (0.92THz) as a function of time delay.

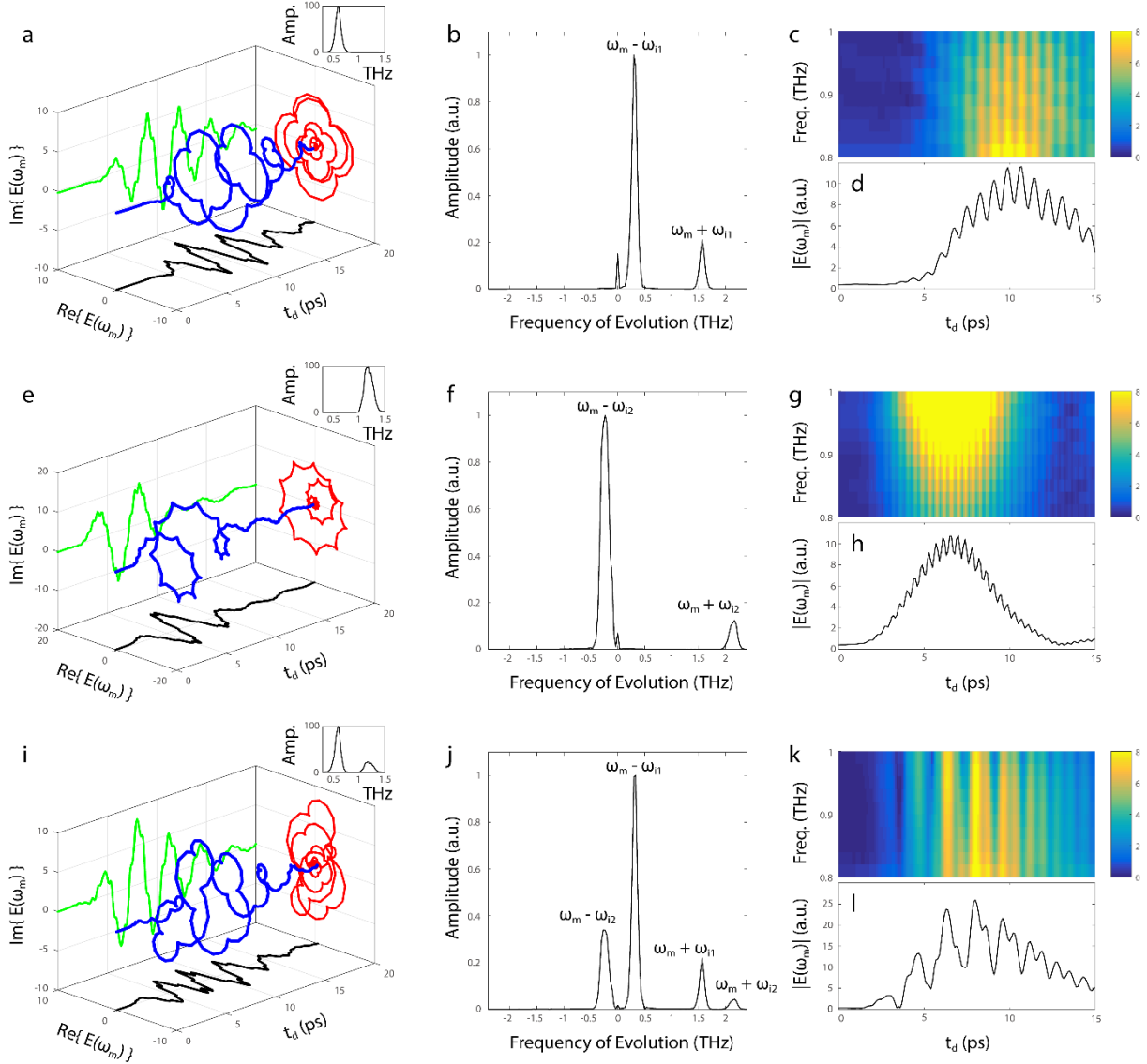

**Figure S12. Calculated complex amplitude of the converted field and its magnitude for multi-cycle input pulses of a lower single band (a-d, with a centre frequency of $\omega_{i1}$), a higher single band (e-h, with a centre frequency of $\omega_{i2}$) and dual bands (i-l, with centre frequencies of $\omega_{i1}$ and $\omega_{i2}$) with FDTD simulation.** (a) Complex amplitude of the converted field at $\omega_m$ plotted as a function of the time delay (blue) and corresponding projections on each of the planes. The inset shows the spectrum of the input pulse. (b) Evolutionary spectrum of the complex amplitude. A Fourier transform has been performed with a respect to the time delay. (c) Magnitude of the complex amplitude for a spectral range of 0.8-1THz. (d) Magnitude of the complex amplitude at $\omega_m$. (e) Complex amplitude of the converted field at $\omega_m$ plotted as a function of the time delay (blue) and corresponding projections on each of the planes. The inset shows the spectrum of the input pulse. (f) Evolutionary spectrum of the complex amplitude. A Fourier transform has been performed with a respect to the time delay. (g) Magnitude of the complex amplitude for a spectral range of 0.8-1THz. (h) Magnitude of the complex amplitude at $\omega_m$. (i) Complex amplitude of the converted field at $\omega_m$ plotted as a function of the time delay (blue) and corresponding projections on each of the planes. The inset shows the spectrum of the input pulse. (j) Evolutionary spectrum of the complex amplitude. A Fourier transform has been performed with a respect to the time delay. (k) Magnitude of the complex amplitude for a spectral range of 0.8-1THz. (l) Magnitude of the complex amplitude at $\omega_m$.